\documentclass[fleqn,10pt]{wlscirep}
\usepackage[utf8]{inputenc}
\usepackage[T1]{fontenc}
\usepackage{bm}

\title{Network Communities of Dynamical Influence}

\author[1,*]{Ruaridh Clark}
\author[2]{Giuliano Punzo}
\author[1]{Malcolm Macdonald}
\affil[1]{Department of Mechanical and Aerospace Engineering, University of Strathclyde, Glasgow, United Kingdom}
\affil[2]{Department of Automatic Control and Systems, University of Sheffield, Sheffield, United Kingdom}

\affil[*]{ruaridh.clark@strath.ac.uk}

\begin{abstract}
Fuelled by a desire for greater connectivity, networked systems now pervade our society at an unprecedented level that will affect it in ways we do not yet understand. In contrast, nature has already developed efficient networks that can instigate rapid response and consensus, when key elements are stimulated. We present a technique for identifying these key elements by investigating the relationships between a system's most dominant eigenvectors. This approach reveals the most effective vertices for leading a network to rapid consensus when stimulated, as well as the communities that form under their dynamical influence. In applying this technique, the effectiveness of starling flocks was found to be due, in part, to the low outdegree of every bird, where increasing the number of outgoing connections can produce a less responsive flock. A larger outdegree also affects the location of the birds with the most influence, where these influentially connected birds become more centrally located and in a poorer position to observe a predator and, hence, instigate an evasion manoeuvre. Finally, the technique was found to be effective in large voxel-wise brain connectomes where subjects can be identified from their influential communities.

\end{abstract}
\begin{document}

\flushbottom
\maketitle

\thispagestyle{empty}

\section*{Introduction}
Human capacity to create networked systems is expanding with the growing popularity of cloud services\cite{idg2018cloud}, improvements in mobile (cellular) network infrastructure and the expansion of the internet, including satellite based provision\cite{henry2019constellation}. Increased connectivity is not without its drawbacks where denial of service attacks, exploiting Internet of Things (IoT) devices, are an example of the vulnerabilities that can emerge in large, complicated, networks\cite{andrea2015internet}. To protect against this growing reliance on networked systems, this paper strives to further our understanding of how network topology influences dynamical response by identifying communities as key pathways of communication from the most influential network nodes.

Artificial networks are sophisticated but they cannot compete with nature's accomplishments, where the human brain is estimated to contain 100 billion neurons and 100 trillion synaptic connections\cite{braun2015human}. 
For the small neuronal system of the {\it Caenorhabditis elegans}\cite{mivsic2014network} and in a low resolution network representation of the macaque connectome\cite{bacik2016flow}, the flow of information has been simulated to reveal the existence of bottlenecks and clusters. Such numerical models of information flow become intractable at a sufficiently large scale, therefore this paper uses a spectral method to uncover communication dynamics in some of nature's most sophisticated networks, including human connectomes with around $850,000$ vertices.

\subsection*{Community Detection}\label{sec:comm_detect}
This paper presents an eigenvector-based community detection method. Most community detection, partitioning and graph clustering methods consider either the edge direction or the graph spectral properties to create network divisions, such as modularity, random walk based methods and spectral clustering\cite{malliaros2013clustering}. 
Modularity detects communities by comparing the density of edges present with the density that would be present if all the edges were reassigned a new destination vertex at random\cite{leicht2008community}. 
Random walk based methods can either emulate the properties of the first eigenvector of the adjacency matrix, see PageRank\cite{page1999pagerank}, or be used to identify communities from modelled diffusion of information\cite{pons2005computing,bacik2016flow}. Two commonly used spectral clustering approaches are normalised cuts\cite{shi2000normalized} and k-means clustering\cite{bradley1998refining}. 
The approach, detailed herein, is most similar to that of k-means clustering whereby multiple eigenvectors of the Laplacian matrix are used to split the graph into communities. In the case of k-means clustering, it employs a heuristic algorithm (k-means) to separate the network into k clusters where each cluster is defined in relation to one of k centroids. These centroids iteratively adjust their positions with the final composition of the clusters dependent on the initial placement of the centroids\cite{bradley1998refining}. There is no optimal solution to the selection of k for any given network instead multiple runs can be performed to determine the sensitivity of k to a chosen criteria\cite{von2007tutorial}. 

In this paper, we detect communities without the use of recursive optimisation, instead the number of communities, their members, and influential vertices are all a product of the network's spectral properties. 
The communities, defined herein, are referred to as Communities of Dynamical Influence (CDI) as each is associated with a vertex that has a high dynamical influence. {\it Dynamical influence} is defined as a measure of how strongly a node's dynamical state affects the collective behaviour (i.e. state) of the network\cite{klemm2012measure}.
These influential vertices are detected with the first left eigenvector of the Laplacian matrix, which has been employed previously to identify the most effective vertices for spreading information or disease in a dynamical network\cite{klemm2012measure}. The first eigenvector of the adjacency matrix is also widely used as a centrality metric\cite{bonacich2007some}, where it identifies vertices that would be visited most often by agents performing a random walk on the graph. Influential communities have been considered before from the perspective of identifying a node grouping that maximises its influence. To achieve these high influence groupings, community detection parameters are varied using for example modularity or $k$-core\cite{seidman1983network}, with each vertex given an influence value that is defined either independently of the network topology, by applying a weighting to the vertices\cite{li2017finding}, or as a product of the topology, by using Katz centrality (a form of eigenvector centrality)\cite{zhan2016identification}. 
Influence can also be assessed after community designation, 
where an influence propagation model can be used to assess each community's influential reach\cite{li2017most}.
In this paper, we do not attempt to maximise the influence of community members but instead consider the influence that propagates from vertices during consensus. Information propagation of linear consensus has been used previously to define influential communities\cite{stanoev2011identifying}, where each community is composed of vertices that are on one of the most effective pathways from an influential vertex. 
In this paper, the same definition is applied for community detection but we do not rely on a model or numerical process, instead communities and dynamical influence can be detected from the relationships between the system's eigenvectors.

This paper includes the application of CDI on a starling flock model and voxel-wise human brain networks. 
Starlings are known to employ a near constant outdegree (six to seven for all flock members\cite{ballerini2008interaction}), which has been found to maximise robustness and allows them to manage uncertainty in consensus\cite{young2013starling}. It has also been observed that starlings on the edge of the flock are those that trigger predator avoidance manoeuvres, as they are first to observe predators\cite{attanasi2015emergence,herbert2015initiation}. The current paper builds on these findings be considering how the outdegree, maintained by starlings, can facilitate their responsiveness to birds attempting to triggers a predator evasion manoeuvre. While the brain analysis compares the influential communities present in the same subject but at different times, using scan-rescan magnetic resonance imaging (MRI) data from Landman et al\cite{landman2011multi}, to not only identify subjects from their neuronal communities but also detect changes in their functional activity.

\subsection*{Influence of Network Perturbations}
To validate the claims that the CDI are the communities with the greatest influence over the consensus process, we investigate the optimisation of consensus leadership. The speed of consensus can be captured analytically by assessing the first eigenvalue of the system matrix\cite{punzo2016using}. Consensus can be driven to a target value by applying a constant input perturbation to a set of vertices that have a directed connection to all other vertices\cite{punzo2016using}. Similar perturbation optimisations have been studied extensively in the context of leadership selection for the control of multi-agent and swarm systems\cite{punzo2016using,fitch2013information,lin2011algorithms,patterson2017optimal,gan2018performance}. In these cases, the perturbation is often constrained so that only a set number of leaders are chosen with a binary option for perturbation input, i.e. vertices set as leaders or followers. In the context of multi-agent systems, there is significantly more literature on minimising the steady-state variance about an input perturbation\cite{fitch2013information,lin2011algorithms} than there is on fast convergence to consensus. There has been an attempt to tackle both problems, but this work was restricted to 1-D community and ring graphs\cite{patterson2017optimal}, and an examination of how the proportion of leader vertices affects the convergence rate to consensus in multiplex networks\cite{gan2018performance}. Of most relevance, to the work herein, is globally bounded input perturbations that can be applied with a variable distribution to any combination of vertices\cite{punzo2016using}. For such a case, the first left eigenvector of the Laplacian matrix was identified as a sub-optimal resource allocation (equivalent to an input perturbation) for achieving fast convergence to consensus\cite{punzo2016using}. An improvement in this allocation has since been developed for directed $k$-outdegree graphs, where a near-optimal perturbation vector can be produced by combining first left eigenvectors from manipulated versions of the adjacency matrix\cite{clark2016consensus}. 
 It is worth noting that the globally bounded perturbation optimisation problem, to maximise convergence rate to consensus, has no verifiable solution. A numerical optimiser can produce near-optimal solutions, but detection of the global minima is not guaranteed. A significant contribution of this article is the filtering out of local minima by highlighting the most effective network influencers, a process that functions at any network scale and is no longer restricted to $k$-outdegree graphs. 

\section*{Results}

\begin{figure}[!tb]
\captionsetup{justification=centering}
\centering
    \includegraphics[width=\textwidth,trim={0mm 45mm 5mm 20mm},clip=true]{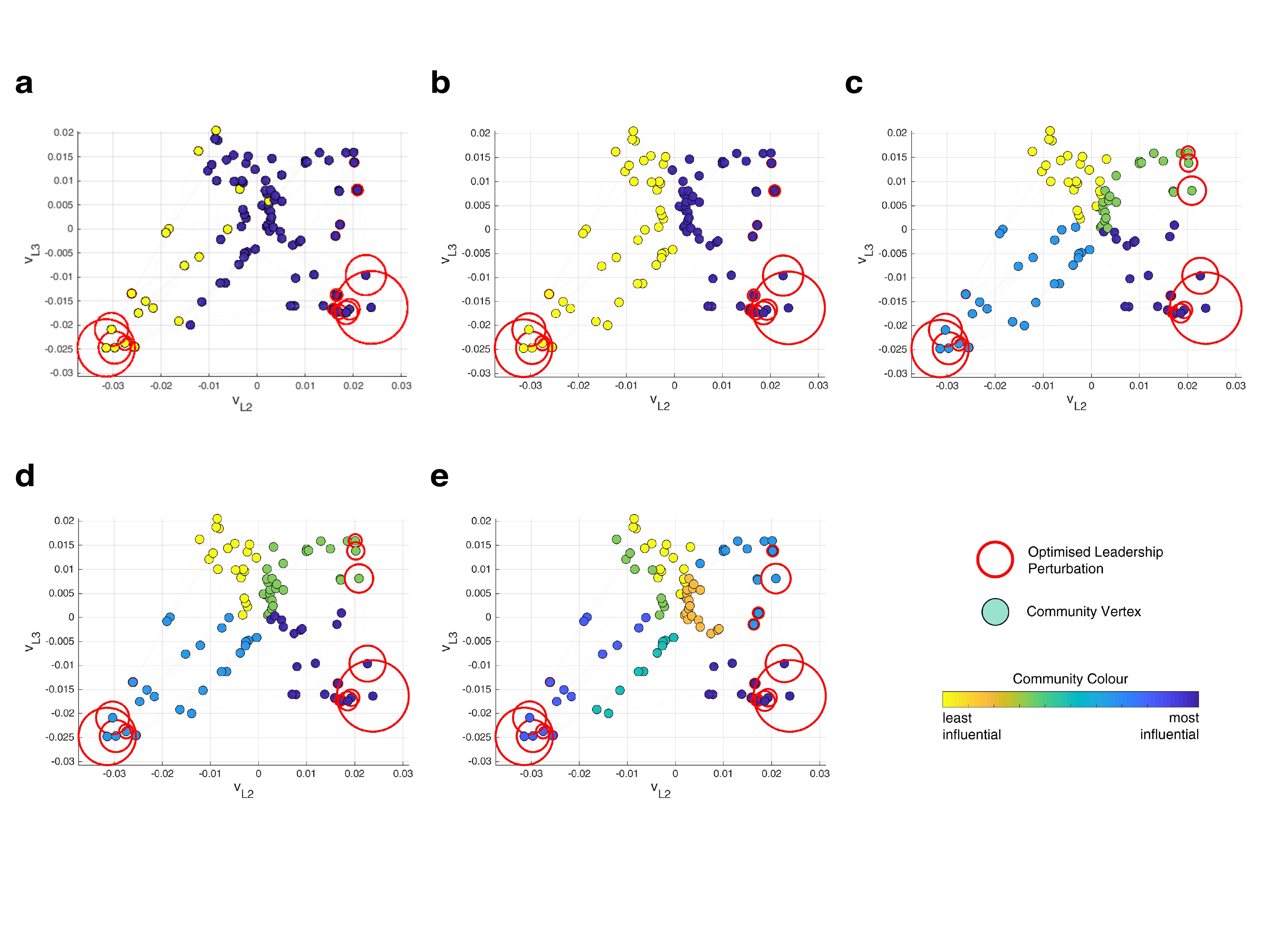} %,height=8cm
  \caption{Communities defined by the CDI method with {\bf a} 1 eigenvector, {\bf b} 2 eigenvectors, {\bf c} 3 eigenvectors, {\bf d} 4 or 5 eigenvectors, {\bf e} 6 eigenvectors. The network is $k$-NNR with 100 vertices, where $k=10$. Communities are denoted by colour according to their influence over the whole network. An optimised perturbation using the chosen community detection overlays the network with circles that are proportional to the perturbation magnitude.}
    \label{fig:CDI_v_CDI}
\end{figure}

The Communities of Dynamical Influence (CDI) are detected using the most dominant left eigenvectors of the Laplacian matrix, as described in the Methods section. 
The CDI are shown in Figure \ref{fig:CDI_v_CDI} for 100 vertex, $k=10$, nearest neighbour ($k$-NNR) networks. The $k$-NNR networks are generated by randomly distributing vertices in a square plane with each vertex connecting to its $k$ nearest neighbours according to euclidean distance. The first left eigenvector (${\bf v}_{L1}$) is always used to determine the dynamical influence of the CDI but, as shown in Figure \ref{fig:CDI_v_CDI} where the axes are ${\bf v}_{L2}$ and ${\bf v}_{L3}$, the other input eigenvectors affect the community composition. This can be seen most clearly in Figure \ref{fig:CDI_v_CDI} {\bf b} where the second eigenvector of the Laplacian matrix (${\bf v}_{L2}$) divides the network in two as CDI only has the first two eigenvectors as input. ${\bf v}_{L1}$ has only positive entries, so community vertices form a trail behind their community leader that ends close to the origin of the eigenvector coordinate system. 

The CDI can be used as an input to a perturbation optimiser, detailed in the Methods section (Algorithm~\ref{Alg:CDI_opt}), that optimises the network's convergence to consensus by maximising the first eigenvalue of the perturbed Laplacian matrix. Essentially, the optimisation identifies the most effective leaders in the network, those with the highest ${\bf v}_{L1}$ values in each community, and applies a perturbation of variable magnitude to those vertices. The leadership perturbations applied from the CDI-based optimisation are detailed in Figure \ref{fig:CDI_v_CDI}, where the 3,4 and 5 eigenvector CDI (Figure \ref{fig:CDI_v_CDI} {\bf b} and {\bf c}) produce the same result and a faster convergence than the 1,2 and 6 eigenvector CDI. The optimal number of eigenvectors, in terms of effective consensus leadership, varies depending on the network. Using only one or two eigenvectors makes it more likely that important community divisions are not identified, as seen in Figure \ref{fig:CDI_v_CDI} {\bf a} and {\bf b}, which produce the slowest convergence by only identifying two communities. Using more eigenvectors can also result in a sub-optimal performance with 6 eigenvectors outperforming the 1 and 2 eigenvector case but not the 3,4 or 5 eigenvector optimisation.

In the following section, three eigenvectors are used for the detection of dynamical influence. CDI with three eigenvectors does not always produce the fastest consensus but in the following section it is shown to produce consistently good results (see Fig.~\ref{fig:opt_k_nnr}). Three eigenvectors provides accurate community division whilst ensuring all communities are amongst the most influential. The issue with using more eigenvectors is that the detected communities may no longer reflect the most effective for leading network consensus. This has already been seen in Figure \ref{fig:CDI_v_CDI} {\bf e} where a community formed (light blue - top right) with vertices that had high ${\bf v}_{L1}$ values because they were connected to more influential vertices in other communities, and not because they were effective network leaders on their own. A perturbation optimisation comparisons between 3 and 4 eigenvector CDI, using the series of $k$-NNR networks from Figure~\ref{fig:opt_k_nnr} {\bf b}, revealed that 4 eigenvector CDI often produced a superior community division, and hence faster convergence, but it was occasionally susceptible to significant inaccuracies. Such inaccuracies were a result of poor community designation for the same reason that Figure \ref{fig:CDI_v_CDI} {\bf e} produced a sub-optimal result.

\subsection*{Validation of Community Influence}
When referring to the influence of a community, we are referring to the influence of the influential vertices in that community. These vertices wield the greatest global influence within their local cluster, but this usually means that they are also the most locally influential vertex. To validate the claim that the CDI are influential, we performed a series of analyses comparing perturbations optimised to drive convergence in linear consensus. The vertices highlighted by this optimisation are those that can lead the network effectively to a new state of consensus, i.e. those with strong local and global network influence, which should align with the influential CDI vertices and their communities.

\begin{figure}[!t]
\captionsetup{justification=centering}
  \centering
    \includegraphics[width=\textwidth,trim={00mm 10mm 0mm 10mm},clip=true]{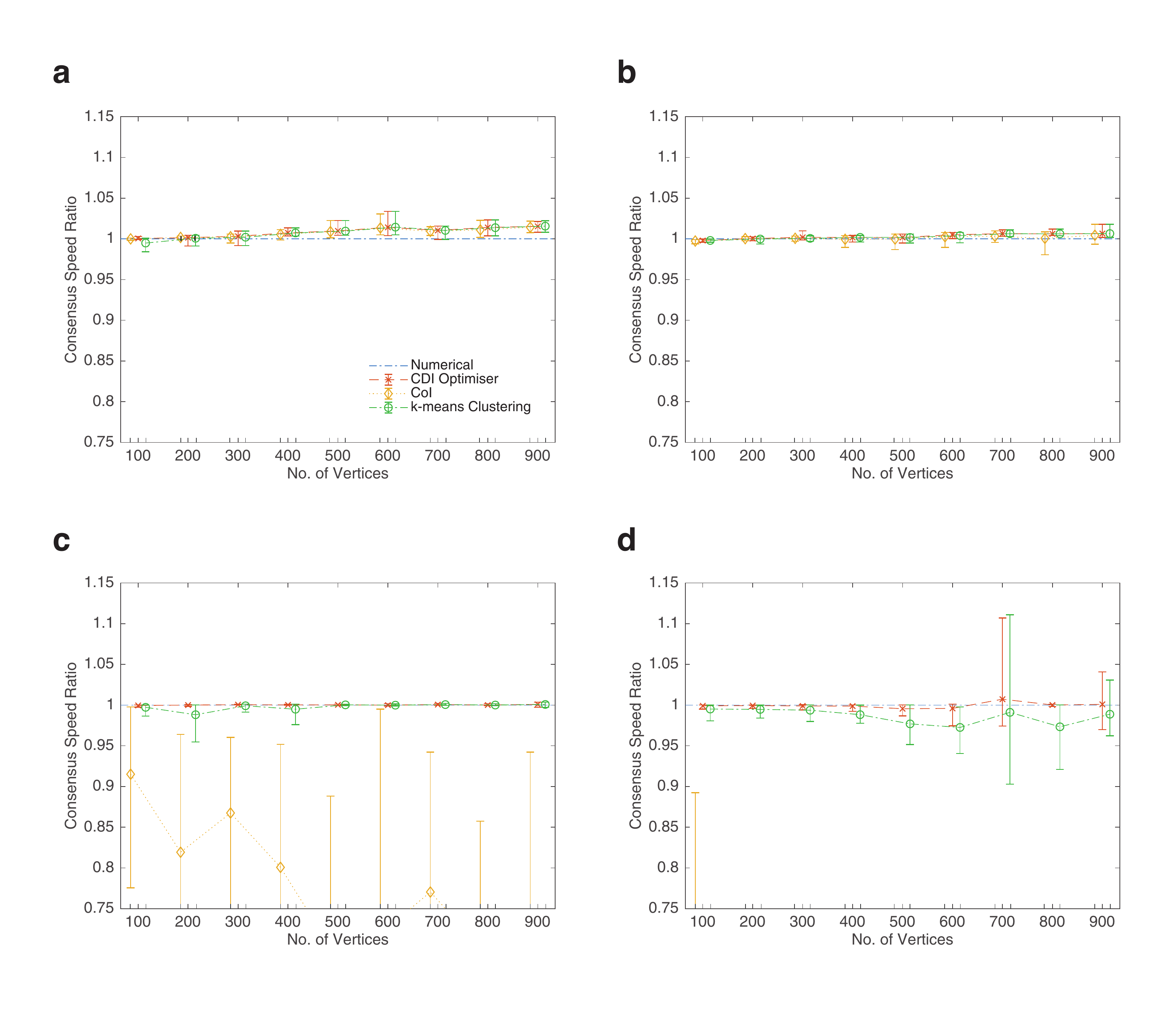} %,height=8cm
  \caption{Consensus Speed Ratio, with reference to a numerical optimiser, for three different optimisations with 10 graphs at each vertex size interval.  Outdegree is set at $k=10$ for {\bf a} Erd\H{o}s-R\'{e}nyi random networks and {\bf b} $k$-NNR networks.  Outdegree is varied between $k=3$ and $k=10$ for each vertex, by sampling at random from a uniform distribution, for {\bf c} Erd\H{o}s-R\'{e}nyi random networks and {\bf d} $k$-NNR networks.}
    \label{fig:opt_k_nnr}
\end{figure}

The CDI and k-means clustering algorithms are used as an input to the perturbation optimiser, detailed in the Methods section (Algorithm~\ref{Alg:CDI_opt}). It should be noted that the k-means clustering requires the number of divisions to be defined and, hence, is set to detect the same number of communities as found by the CDI algorithm. The results of these optimisations are compared with the Communities of Influence (CoI) method\cite{clark2016consensus} and a numerical optimiser\cite{mathworks2015constrained} in Fig.~\ref{fig:opt_k_nnr}. 
The CoI method generates optimised perturbations by detecting influence, using the first left eigenvector, and investigating how this influence changes when key vertices are removed from the network. This was shown to be effective in $k$-outdegree networks where the CoI method, using 5 input vectors, produced similar results to the output of a numerical optimiser\cite{clark2016consensus}. In Fig.~\ref{fig:opt_k_nnr}, the CDI optimiser is shown to produce similar result in $k$-outdegree networks to the CoI method, but achieves notably superior results when applied to variable $k$ networks. In fact the CoI results are too low to be included for many of the networks sizes reported in Fig.~\ref{fig:opt_k_nnr} {\bf c} and {\bf d}. 
The CDI-based optimiser frequently exceeds the results of the numerical optimiser, which struggles to find globally optimal minima.

The k-means clustering algorithm developed by Ng et al.\cite{ng2002spectral} also employs eigenvectors and a form of machine learning to define clusters. 
Fig.~\ref{fig:opt_k_nnr} shows that the k-means clustering algorithm, when provided with the number of communities found by the CDI algorithm, can often identify the most effective network leaders by placing them in separate communities. The main differences between CDI and k-means clustering are seen in the variable outdegree $k$-NNR case, Fig.~\ref{fig:opt_k_nnr} {\bf d}, where the CDI performs consistently better.

\begin{figure}[!b]
\centering
\captionsetup{justification=centering}
\includegraphics[width=.75\textwidth,trim={10mm 40mm 109mm 20mm},clip=true]{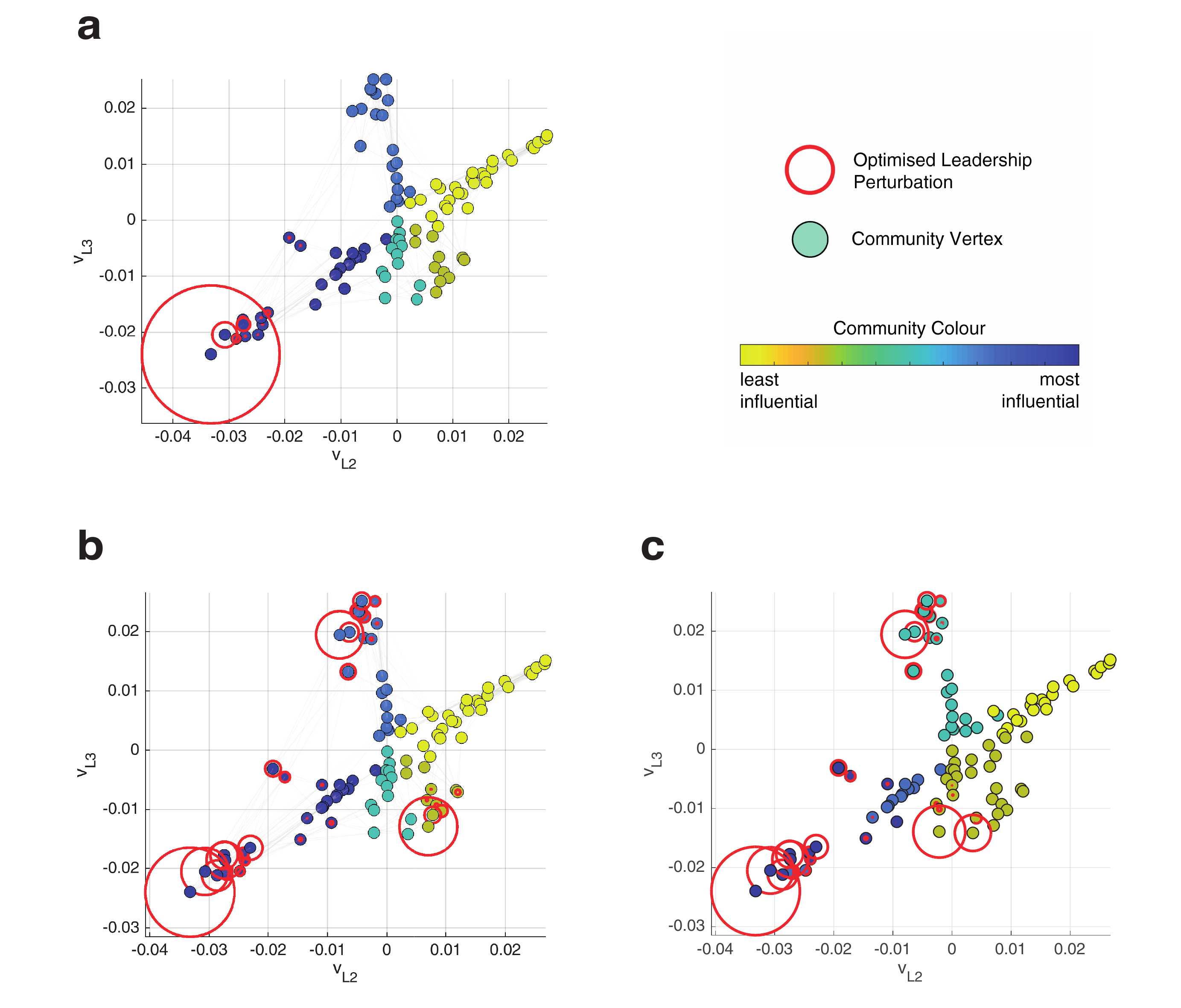} 
  \caption{Communities defined by the CDI method with 3 eigenvectors for {\bf a} edge weights of 1, and {\bf b} edge weights of 0.2 and in {\bf c} k-means clustering is applied for edge weights of 0.2. The network is $k$-NNR with 100 vertices, where the outdegree varies between 3 and 10 by sampling at random from a uniform distribution. Communities are denoted by colour according to their influence over the whole network. An optimised perturbation using the chosen community detection overlays the network with circles that are proportional to the perturbation magnitude.} %$0.00275$
\label{fig:opt_CDI_SC}
\end{figure}

The superiority of CDI in comparison with k-means clustering is made clearer by reducing the weight of each edge for a given graph. This alteration constricts the flow of information through the graph, with the result that perturbations need to be spread to a greater number of vertices to overcome this restriction. This phenomena is seen in Fig.~\ref{fig:opt_CDI_SC} {\bf a} and {\bf b} for a 100 vertex $k$-NNR topology with edge weight set at $1$ and $0.2$ respectively. If the edge weight was reduced to 0, then the optimal perturbation would provide all vertices with the same perturbation, i.e. lead them all individually. By requiring more network leaders, the difference between CDI and k-means clustering can be seen even for the $k$-NNR topology, where the methods performed similarly in Figure~\ref{fig:opt_k_nnr} {\bf b}. Comparing CDI in Fig.~\ref{fig:opt_CDI_SC} {\bf b} with k-means clustering in Fig.~\ref{fig:opt_CDI_SC} {\bf c} shows that the clusters generated by both methods are similar. But in the bottom right quadrant of the figures it can be seen that CDI detects a division that k-means does not and this results in a faster convergence speed for the CDI-based optimisation. These small differences in community designation are also what differentiates the methods in Fig.~\ref{fig:opt_k_nnr} {\bf d}.

\subsection*{Responsive Starling Flock Topology}\label{sec:starling}
\begin{figure}[!tb]
\centering
\captionsetup{justification=centering}
\includegraphics[width=\textwidth,trim={20mm 10mm 20mm 0mm},clip=true]{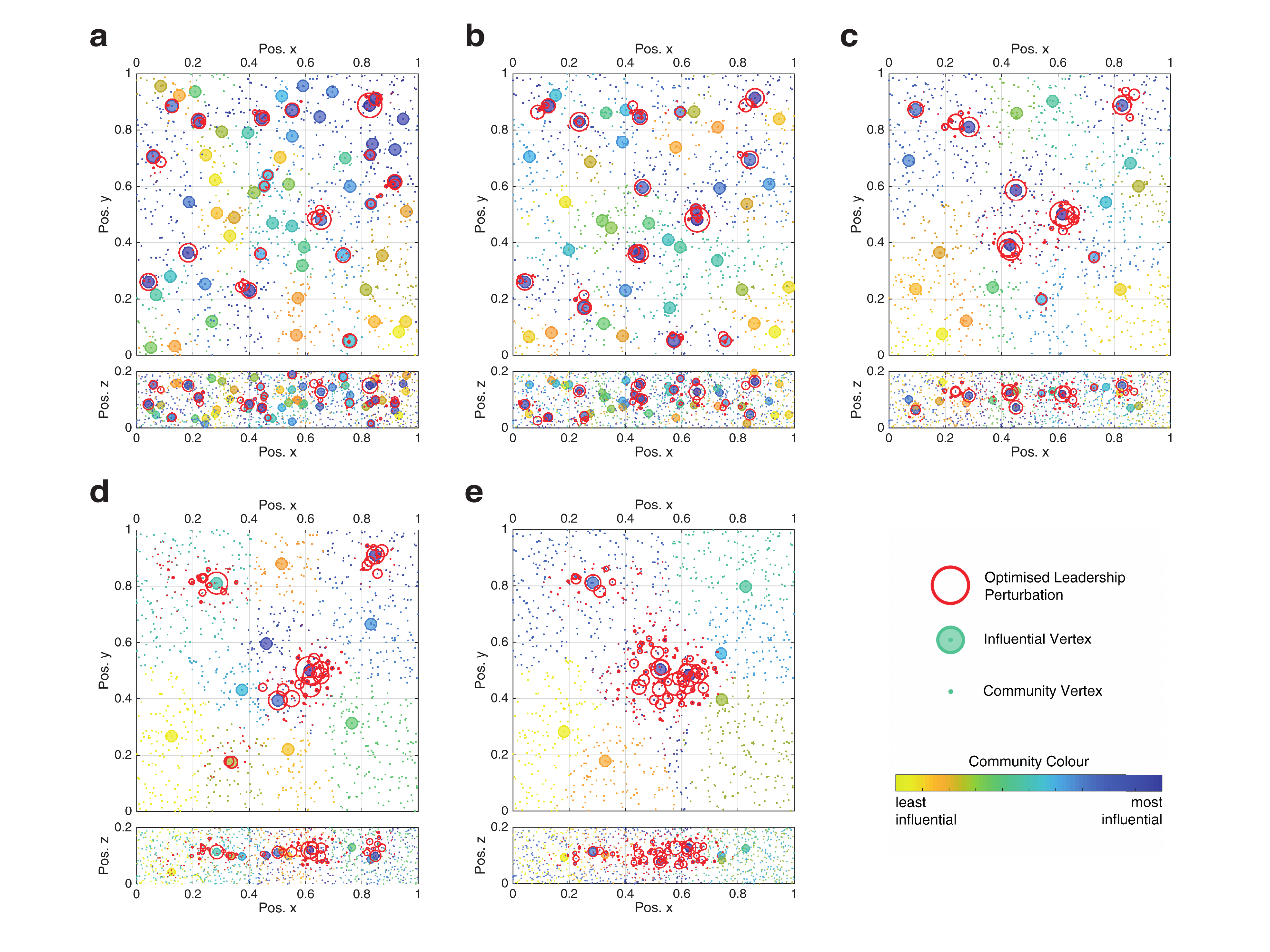}
  \caption{Starling communities are denoted by colour according to the CDI method with the three most dominant eigenvectors for a 1200 vertex $k$-NNR starling flock model where {\bf a} $k=5$, {\bf b} $k=7$, {\bf c} $k=15$, {\bf d} $k=25$ and {\bf e} $k=50$. A single influential vertex is displayed for each community with a shaded circle and the optimised leadership perturbation using 3 eigenvector CDI is denoted with circles that are proportional to the perturbation magnitude.}
\label{fig:starling}
\vspace{-4mm}
\end{figure}

Starling flocks tend towards a thicknesses of between 0.13 and 0.27, where the flock thickness is the ratio of the smallest to largest dimension of an ellipsoid having the same principal moments of inertia as the flock\cite{young2013starling}. In Fig.~\ref{fig:starling}, five examples of 1200 bird starling flock networks are presented with a thickness of 0.2, where the flock is modelled by randomly distributing birds from a uniform distribution within a rectangular prism\cite{young2013starling}.
The community distribution and optimised perturbations are shown for these representative examples that employ $k$-NNR topologies for the following outdegrees; $k=5$, $7$, $15$, $25$ and $50$. The starling vertices remain in the same position for each analysis with the number of nearest neighbours, $k$, the only change that affects the network structure. 

What is of particular interest, from these results, is the position of the influential vertices in each example. In high $k$ outdegree examples, such as Fig.~\ref{fig:starling} {\bf c}, {\bf d} and {\bf e}, the influential vertices, associated with the most influential communities, are centrally located. An optimised perturbation (using  3 eigenvector CDI) is shown to align with the most influential communities, therefore the perturbations become increasingly centrally located as the outdegree increases. For the $k=50$ example, in Fig.~\ref{fig:starling} {\bf e}, these perturbations are primarily located in the centre of the model flock. For lower outdegree examples, such as Fig.~\ref{fig:starling} {\bf a} and {\bf b}, the influential vertices, especially those associated with the most influential communities, are more evenly distributed throughout the flock, aided by the increase in the number of CDI present. Considering an actual starling flock where $k\approx 7$, the topology makes it more likely, in comparison with a higher outdegree scenario, that one of the most influential birds will be near the site of a predator attack. Therefore, the low outdegree topology is more likely to have a highly influential bird involved in leading a predation avoidance manoeuvre.

Varying the outdegree for the starling flock model, while using the same distribution of 1200 birds, and applying an optimised leadership perturbation changes the value of the dominant eigenvalue $\lambda_1$ of the perturbed Laplacian matrix, which represents convergence speed.  For outdegrees between $k=5$ and $k=50$, $\lambda_1$ results roughly conform ($R^2 = 0.962$) to a power law distribution ($\lambda_1 = 0.0031 k^{-0.184}$). A higher convergence speed is indicative of a graph that can be more effectively led by key vertices. The highest convergence speeds, therefore, usually belong to networks with lower outdegrees.
Given that a lower outdegree tends to result in faster perturbation driven consensus, the reason for the outdegree remaining as high as 7 may be due to the requirements of maintaining a connected flock. If the outdegree is too low the flock may split whenever a perturbation is applied and may not reconnect. Defining a lower bound that ensures connectivity is beyond the scope of this paper, but lower bounds have been identified for static topologies including $k$-NNR graphs where the required $k$ increases with the number of vertices\cite{balister2005connectivity}.

\subsection*{Brain Similarity Detection}
The CDI method is applied in this section on large-scale human connectomes generated by Roncal et al. \cite{roncal2013migraine}
from magnetic resonance imaging (MRI) scans carried out by Landman et al.\cite{landman2011multi}. Landman et al. used 21 healthy volunteers, where each subject was scanned twice with a short break between scan and rescan (scan 1 and scan 2 respectively)\cite{landman2011multi}. Note that one of the scans, for {\it subject 127}, could not be sourced and, hence, this article shall consider the remaining 20 volunteers.

The networks generated by Roncal et al. are undirected and the adjacency matrix, rather than the Laplacian is used, due to the scale of the network (see Methods section). Therefore, the CDI no longer identifies the most effective leaders of system consensus as every vertex both leads and follows equally. This makes it impossible to differentiate between important sources and sinks of information in the network. The use of the adjacency matrix also means that the communities are ranked by popularity, i.e. the frequency with which a random walker would visit each vertex in the graph, rather than effective consensus leadership. Therefore, instead of drawing conclusions about a community's network leadership, the CDI is used as a similarity metric by detecting changes in the influential communities (whether sources or sinks) of brain connectomes.

\begin{figure}[!t]
  \centering
  \captionsetup{justification=centering}
         \includegraphics[width=.95\textwidth,trim={0mm 17mm 0mm 10mm},clip=true]{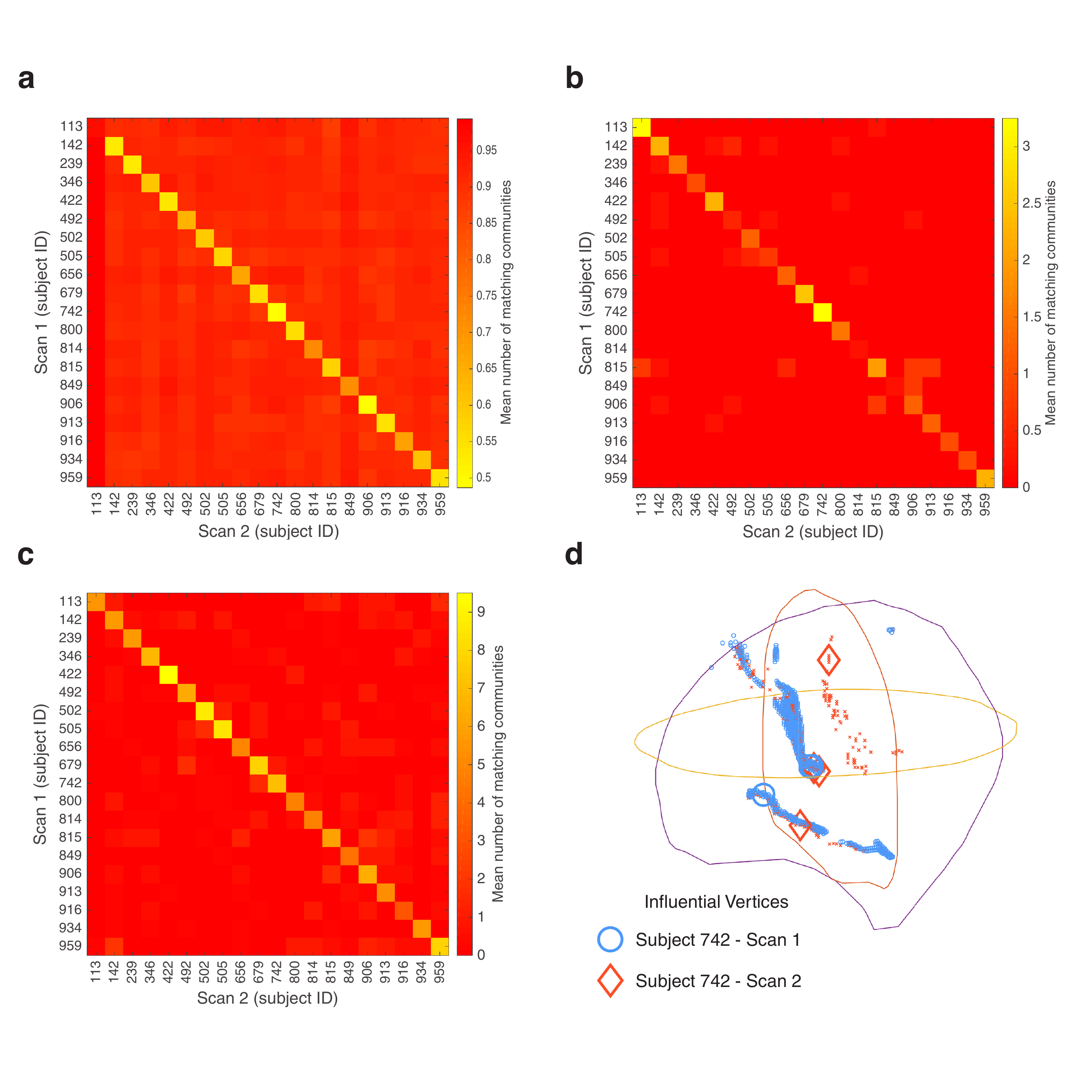}
    \caption{Comparison of scan 1 and scan 2 from Landman et al.\cite{landman2011multi} subjects using  {\bf a} {\it edge graph edit distance}, and the {\it mean number of matching communities} with CDI using {\bf b} 3 eigenvectors and {\bf c} 10 eigenvectors of the adjacency matrix. {\bf d} displays the communities detected by CDI, using 3 eigenvectors, for scan 1 and 2 of subject 113 where the vertices are shown on a left lateral angled view of a brain, modelled with outlines from the x, y and z planes.}\label{fig:ALL_K} 
\end{figure}

In the work by Roncal et al.\cite{roncal2013migraine}, the Frobenius Norm was successfully used to detect the similarity of these scan-rescan matrices created from Landman et al.'s study\cite{landman2011multi}. The Frobenius Norm is an established matrix distance measure\cite{zuo2006assembled}, referred to as Frobenius Distance when assessing graph similarity. The result from Roncal et al.’s study\cite{roncal2013migraine} has not proven to be exactly reproducible with the published dataset\cite{landman2011multi},
as
the scan-rescan comparisons do not always produce the lowest values (i.e. greatest similarity).
A superior similarity metric for this case was identified as Graph Edit Distance (GED), which is an inexact graph matching method defined as the cost of the least expensive sequence of edit operations that are needed to transform one graph into another \cite{robles2005graph}. Specifically, the results of
{\it edge GED}\cite{manrique2018comparing} are displayed in Figure~\ref{fig:ALL_K} {\bf a} for the scan 1 and 2 (scan and rescan) comparisons. GED has been used as an identification method for matching fingerprints\cite{robles2005graph}, but it fails to identify subject 113 from their scan-rescan comparison. The CDI based comparisons are shown in Figure~\ref{fig:ALL_K} {\bf c} and {\bf d} for three and ten input eigenvectors respectively. The CDI communities with 3 input eigenvectors are also displayed in three-dimensional space (Figure~\ref{fig:ALL_K} {\bf d}) for subject 113, with the difference in community density providing an insight into why edge GED failed but CDI succeeded in recognising subject 113. The paths taken by communities in Figure~\ref{fig:ALL_K} {\bf d} are mostly similar with one non-matching community present from scan 2. However, the density of these communities are significantly different with far more vertices in the scan 1 communities than those from scan 2. This also translates to the density of the connectivity network, where scan 1 has 841,097 vertices with a non-zero outdegree and scan 2 only has 647,049 vertices for subject 113. This difference of 194,048 vertices is far larger than any other scan-rescan comparison, where the next largest difference in 25,554 vertices (mean number of non-zero outdegree vertices is 836,699 for the other scans). This difference in graph density appears to prevent edge GED from detecting a match.

The CDI community matching, with ten input eigenvectors (Fig.~\ref{fig:ALL_K} {\bf c}), presents a similar accuracy to edge GED (Fig.~\ref{fig:ALL_K} {\bf a}). 
But it is notable that 3 eigenvector CDI produces some poor matches in Fig.~\ref{fig:ALL_K} {\bf c}, such as subject 814 and 849. This suggests that the most influential communities have changed between scan and rescan for these subjects as they are still recognisable when considering the less influential communities detected when using ten eigenvectors. 

Similar results to Fig.~\ref{fig:ALL_K} {\bf c} can be obtained by taking an approach based on normalised cut\cite{shi2000normalized}, where the Fiedler vector divides the graph by the sign of the vector's entries. This spectral bisection approach, as described in the Methods section, does not manage to clearly identify all subjects when applying the {\it mean number of matching communities} procedure. It also creates more off-diagonal false matches than Figure~\ref{fig:ALL_K} {\bf c} and does it give insight into any changes in neuronal influence.

\section*{Discussion}

In this article a method for detecting Communities of Dynamical Influence (CDI) is proposed that uses the relationship between a selection of the most dominant left eigenvectors of the Laplacian to divide a network. The communities are shown to be lead by the most influential vertices when comparing with a perturbation optimiser that maximises the network convergence rate to consensus. CDI defines community divisions that can be similar to those detected by k-means clustering, but user selection is required to define k. In contrast, with CDI the number of community divisions are a product of the network topology and the number of input eigenvectors. Three input eigenvectors are shown to produce consistently good results when using CDI-based consensus optimisation, which is used herein as an assessment of the quality of the influential communities that are found. The selected number of input eigenvectors is a trade-off, for CDI, as more eigenvectors may highlight more or new communities but those communities
may no longer represent the communities that form through association with an influential leader. 
There is scope to consider in future work weighting the contribution of less dominant eigenvectors to ensure high accuracy community division that is still focused on the most influential leaders.

For a starling flock model, the CDI reveal the benefit to starlings of maintaining a low outdegree. Higher outdegrees were seen to reduce the responsiveness of the network, in the CDI-based perturbation optimisations, with the flock also becoming composed of fewer CDI. It was also seen that the most influential vertices became more centrally located within the flock, where they are unlikely to detect an incoming predator. The CDI did not reveal any optimality in the chosen starling outdegree, instead it was suggested $k=7$ may have emerged as a compromise between ensured connectivity and fast response.

In this article a series of human brain networks, around $850,000$ vertices in size, were analysed to identify neuronal communities. The identified communities enabled separate MRI scans to be clearly recognised as belonging to the same subject, especially when using CDI with ten input eigenvectors. We conjecture that the subjects with the lowest mean number of matching communities, when using CDI with the first three eigenvectors, are those that display the greatest change in neuronal activity. Since these same subjects are clearly identified when considering the less influential communities detected by the 10 eigenvector CDI. Edge Graph Edit Distance is shown to be highly effective in identifying the scan similarity for the majority of matches with one exception. This exception highlights the effectiveness of the CDI approach, where pathways/community similarity was still clear even when comparing graphs and communities of significantly different sizes.

\section*{Methods} \label{sec:methods}

A graph is defined as $\mathnormal{G=(V,E)}$, where there is a set of $\mathnormal{V}$ vertices and $\mathnormal{E}$ edges, which are unordered pairs of elements of $\mathnormal{V}$ for an undirected graph and ordered pairs for a directed graph. 

The adjacency matrix, $A$, is a square $n \times n$ matrix when representing a graph of $n$ vertices. This matrix captures the network's connections where $a_{ij}>0$ ($a_{ij}$ is the $ij$\textsuperscript{th} entry of the graph's adjacency matrix) if there exists a directed edge from vertex $i$ to $j$ and 0 otherwise. Variable edge weights contain information on the relative strength of interactions, whilst uniform edge weighting either only represents the presence of a connection or is a result of all the edges having the same information carrying capacity. For an undirected graph, the adjacency matrix is symmetric with an edge $(i,j)\in \mathnormal{E}$ resulting in $a_{ij}=a_{ji}>0$. For a directed graph, the indegree is equal to the column sum, $\sum_i a_{ij}$, and outdegree is equal to the row sum, $\sum_j a_{ij}$.

The Laplacian matrix is defined as
$
L=D-A
$
%S
 where the degree matrix, $D$, is a diagonal matrix and the $i$\textsuperscript{th} diagonal element is equal to the outdegree of vertex $i$. 
The first eigenvalue of the adjacency matrix ($\lambda_1$), referred to as the {\it Spectral Radius}, is the largest eigenvalue in magnitude and is associated with the {\it Perron vector}, which is an eigenvector that contains only positive entries. Whereas for the Laplacian matrix the first eigenvalue is associated with the eigenvalue $\lambda_1 = 0$\cite{punzo2016using}. For a directed graph, the left eigenvectors of the Laplacian matrix, ${\bf v}_{L}$, are row vectors satisfying $ {\bf v}_{L}L =\lambda{\bf v}_{L}$.

\subsection*{Communities of Dynamical Influence}\label{sec:CDI} 
The Communities of Dynamical Influence (CDI) are found by analysing the left eigenvectors of the Laplacian matrix as presented in Algorithm~\ref{Alg:CDI}. The algorithm defines a coordinate system using only the Real part of the selected eigenvectors for a chosen number of input eigenvectors. Hence, if any of the eigenvectors considered form a complex conjugate pair then the algorithm will only ever use one vector from the pair and then choose the next dominant eigenvector not in the pair, since the real part of the complex conjugates will be identical. 

The CDI are based on first identifying the most effective community leaders. These vertices do not "follow" nodes with greater community influence, therefore they are identified as those with no outward connections to a vertex that is further from the origin in this eigenvector-based coordinate system. 

A vertex is assigned to a community when there is a directed path from that vertex to a community leader, with each vertex on the path further from the origin of the eigenvector coordinate system than the previous.
 The number of communities is equal to the number of community leaders where a vertex can belong only to one community. If a vertex is assigned to multiple communities, then it is kept in the community where it is most aligned with the community leader and removed from the others. This alignment is determined by comparing the position vector of the vertex with respect to those of the most influential vertices using the scalar product.

\begin{algorithm}[ht]
	\caption{Detecting Communities of Dynamical Influence (CDI)}
	\label{Alg:CDI}
	\begin{algorithmic}[1]\raggedright
	    \State Set $y$ as the number of input eigenvectors
    	\State ${\bf v}_{Li}$ is the $i$\textsuperscript{th} eigenvector of the normalised Laplacian matrix , $L \in {\rm I\!R}^{n\times n}$
        \State $\bm{v}_1 \gets {\rm I\!R}[{\bf v}_{L1}]$, $k \gets 1$, $j \gets 2$
        \While {$k<y$} 
            \If {${\rm I\!R}[{\bf v}_{Lj}] \neq \bm{v}_t ~\forall ~t=1,...,k$}
                \State $k \gets k+1$
                \State $\bm{v}_k \gets {\bf v}_{Lj}$ (i.e. include only one eigenvector from any complex conjugate pair)
            \EndIf
            \State $j = j + 1$
        \EndWhile
        \State $\bm{e} \gets [\bm{v}_1,..,\bm{v}_y]$, ${\bf s}\gets |\bm{e}|$, $h \gets 1$
        \For {$j \gets 1 ~{\bf to} ~n$}
            \State $\bm{o}$ is the set of $k$ vertices connected to vertex $j$ where $L(j,\bm{o}[t])<0 ~\forall ~t \gets 1$ to $k$
            \If{$\bm{o} \gets \emptyset$ (empty set)}
                \State $s_o\gets0$
            \EndIf
            \If{$s_j > s_{\bm{o}[t]} \forall ~t \gets 1$ {\bf to} $k$}
                \State $\beta_h \gets j$ is a community leader vertex
                \State $h \gets h + 1$
            \EndIf
        \EndFor
        \For {$j \gets 1 ~{\bf to} ~h$}
            \State ${\bf c}_i$ is a list of $d$ vertices that lie on a directed path leading to $\beta_j$ and composed of edges where $s$ 
            \Statex \hspace{5mm}increases, i.e. ${s}_{\it source}<{s}_{\it destination}$
        \EndFor
        \For {$j \gets 1 ~{\bf to} ~n$}
            \State $z_k\gets({e}_{\beta_k} \bigcdot {e}_{j}) / s_{\beta_k}$ $\forall ~k$ where $j \in {\bf c}_k$ (i.e. scalar projection of vertex $j$ on leader vertices ${\beta_k}$)
            \State Remove vertex $j$ from ${\bf c}_{k}$ $\forall ~k$ where $z_k \neq \max_k z_k$
    	\EndFor
    	\State ${\bf c}'\gets {\bf c}_i$ where $i$ is a an ordered list of communities, descending order, according to their largest $\bm{v}_1$ value, eg. $j \in {\bf c}'_1$ where $\bm{v}_1[j] = \max \bm{v}_1$
	\end{algorithmic}
\end{algorithm} 

\subsection*{Perturbation Driven Consensus}\label{sec:pert_cons}
The networks considered herein have {\it n} agents connected via local communication with a static, time-invariant, topology. 
A uniform signal ${\bf u}={\rm{u}[1,1,\cdots,1]}^\intercal{\in {\rm I\!R}^n}$ is supplied to all agents with different positive gains $c_i$, where $i = 1,2,...,n$. The dynamics of this system are defined as
\begin{equation}
\label{eq:dynamics3}
 \dot{x}_i = \sum_{j=1}^n a_{ij} (x_j - x_i) + c_i ({u} - x_i)
\end{equation}
where $x_i$ is the state of the $i^{\text{th}}$ agent and $\it{u}$ is the scalar target value that all agents must achieve. The resource allocation, $c_i$, ranges from 0 to 1, is globally bounded as ${\sum_i c_i = 1}$, and scales the comparison between the uniform input signal, $u$, and the current state $x_i$.

 The global dynamics of the network can be expressed with respect to the Laplacian matrix as
\begin{equation}
\label{eq:dynamics_global}
\dot{\bf x} = - L{\bf x} + C({\bf u}-{\bf x})
\end{equation}
where $C$ is the perturbation matrix, $C \gets \text{diag}(\bm{c}) = \text{diag}(c_1,...,c_\mathnormal{P})$. Spanning trees have been highlighted previously as a condition for consensus \cite{shao2015relative, jadbabaie2003coordination, mesbahi2010graph} since for a directed network $\mathnormal{G}$, defined by Eq.~\eqref{eq:dynamics_global}, consensus will eventually be achieved if all agents are reachable, via directed edges, from the vertices supplied with perturbation input.

\subsection*{Perturbation Optimisation}
A perturbation optimisation is presented as a method for validating the CDI algorithm's ability to identify the most influential communities and network leaders. The objective function of this optimisation is to maximise the system's convergence to consensus, by applying a globally bounded perturbation to the vertices. It has been demonstrated that, by changing the coordinates, Eq.~\ref{eq:dynamics_global} can be written as
\begin{equation}
\frac{d{\bf y}}{dt} = -(L+C){\bf y}.
\end{equation}
where the diagonal elements of $C$ can be optimised to maximise the magnitude of $\lambda_1(-(L+C))$, which is the 
most dominant (rightmost) eigenvalue (i.e. eigenvalue with the largest real part) of the negated and perturbed Laplacian matrix\cite{punzo2016using}.

The first step is to optimise a perturbation only applied to most influential vertex from each community, according to their ${\bf v}_{L1}$ value, as detailed in Algorithm~\ref{Alg:leader_opt}. If the optimiser reduces the perturbation applied to the influential vertex to less than or equal to zero then the community is discarded from the optimisation. Once the optimiser has converged, only the communities associated with influential vertices that still have positive perturbations are included in the next step of the optimisation.

\begin{algorithm}[ht]
\caption{Community Leader Optimisation}
\label{Alg:leader_opt}
\begin{algorithmic}[1]\raggedright
        \State Set $\bm{v}_1 \gets {\rm I\!R}[{\bf v}_{L1}]$
		\State Detect $m$ ordered Communities (${\bf c}'_i$) using Algorithm \ref{Alg:CDI}.
		\State ${\bf c}'_i \subseteq \mathnormal{V}$ is the set of vertices belonging to community $i$, where $i\gets 1$ {\bf to} $m$
		\For {$i \gets 1$ {\bf to} $m$}
		    \State $\epsilon _i \gets \max_t \bm{v}_1[t] ~\forall ~t \in {\bf c}'_i$
		    \State $\beta_i \gets j$ where $\bm{v}_1[j] = \max_t \bm{v}_1[t]$
		\EndFor
		\State Sort $\epsilon$ in descending order, ${\epsilon}' \gets \{\epsilon_{r_1},\epsilon_{r_2},...,\epsilon_{r_m}\}$, where $\bm{r}$ is the ordered index vector such that $\epsilon_{r_1}= \max_t \bm{v}_1[t] ~\forall ~t \in \mathnormal{V}$
        \State Set perturbation vector ${\bf p}[1, ..., n] \gets 0$ where $n$ is the number of vertices ($\mathnormal{V}$)
        \For {$i \gets 1$ {\bf to} $m$}
            \State ${\bf p}[\beta_i]\gets\frac{1}{i}$
        \EndFor
		\State Maximise $|\lambda_1(-(L+C))|$ where $C=\text{diag}({\bf p})$, by adjusting ${\bf p} ~\forall ~ p_i > 0$ where {$i \gets 1$ {\bf to} $n$}
		\State If ${\bf p}[\beta_i]\le 0 ~\forall$ {$i \gets 1$ {\bf to} $m$}, set ${\bf p}[\beta_i]\gets 0$ and repeat Step 13, until optimiser convergence.
%		\EndProcedure
\end{algorithmic}
\end{algorithm} 

For each of the selected CDI from Algorithm~\ref{Alg:leader_opt}, an input vector $\bm{\omega}_i$ is created for each CDI with entries populated with their ${\bf v}_{L1}$ entries if the vertex is in the community and values set to zero otherwise. These vectors are then manipulated, using the {\it Power Optimisation} method\cite{clark2016consensus} in Algorithm~\ref{eq:power}, and combined to produce the final optimised perturbation, with weighting variables used to determine the ratio of each vector's contribution in Algorithm~\ref{Alg:CDI_opt}. 

The {Power Optimisation} focuses resources on the most effective leaders by raising an eigenvector to a power, $\eta_i$, for a given input vector, $\bm{\omega}_i$, according to 
\begin{equation}
\label{eq:power}
{\bf p}_i=\frac{\bm{\omega}_i^{\eta_i}}{\sum{({{\omega}_i}^{\eta_i}})}
\end{equation}
where $\bm{\omega}_i^{\eta_i}$ indicates an element-wise operation and the denominator ensures that $\sum{({{\omega}_i}^{\eta_i})}$. When $\eta\rightarrow 0$ the vector, ${\bf p}_i$, approaches a uniform vector state. As $\eta$ is increased, the {Power Optimisation} method iteratively reduces the value of the smaller vector elements while increasing the value of the larger elements. 

In the following equation $t$ power optimised input vectors, ${\bf p}_i$, are combined using weighting variables, ${\bm{r} = \{r_1,...,r_{t}\}}$, to produce the optimised perturbation vector as follows 
\begin{equation}\label{eq_cdi}
	\bm{c}=\frac{\sum_{j=1}^{t}\frac{{\bf p}_{j}}{r_j}} { \sum_{j=1}^{t}\frac{1}{r_j}}
\end{equation}
where the denominator ensures that $\sum_j ({c})_j = 1$. Also note that $C=\text{diag}(\bm{c})$ in the $-(L+C)$ system.

Algorithm~\ref{Alg:CDI_opt} presents the perturbation optimisation procedure, where Eq.~\ref{eq:power} and ~\ref{eq_cdi} are used repeatedly with different inputs and constraints. 
\begin{algorithm}[!ht]
\caption{Perturbation optimisation using CDI}
\label{Alg:CDI_opt}
\begin{algorithmic}[1]\raggedright
%	\Procedure{Globally Bounded Perturbation Optimisation}{}
    \State Input $m$ CDI communities (${\bf c}'$), perturbation vector (${\bf p}$), leader vertex list ($\bm{\beta}$) from Algorithm~\ref{Alg:leader_opt}, $h\gets 1$, $n$ is the number of vertices ($\mathnormal{V}$), $\bm{v}_1 \gets {\rm I\!R}[{\bf v}_{L1}]$
    \For{$j \gets 1$ {\bf to} $m$}
        \State $\bm{\omega}_h[1,...,n] \gets 0$
        \If{${\bf p}[\beta_j]>0$}
            \State $\bm{\omega}_h \gets \bm{v}_1[i] ~\forall ~i \in {\bf c}'_j$ (i.e. communities to be included in the optimisation)
            \State $h\gets h + 1$
        \EndIf
    \EndFor
	\State Maximise $|\lambda_1(-(L+C))|$, where $C\gets\text{diag}({\bf p})$, by optimising $\eta_1$ in Eq.~\ref{eq:power}
	\State Set $\Lambda \gets |\lambda_1(-(L+C))|$, ${\bm{r} \gets \{1\}}$, ${\bm{\eta} \gets \{\eta_1\}}$, $\bm{\omega}\gets\{\bm{\omega}_1\}$
	\For {$j \gets 2$ {\bf to} $h$}
	    \State $q$ is the current number of input vectors ($\bm{\omega}$), ${\bm{r} \gets \{\bm{r},\bm{r}[q-1]\}}$, ${\bm{\eta} \gets \{\bm{\eta},\bm{\eta}[q-1]\}}$, $\bm{\omega}\gets\{\bm{\omega},\bm{\omega}_j\}$, ${\bm{k} \gets \bm{r}}$
        \State Maximise $|\lambda_1(-(L+C))|$, where $C \gets \text{diag}(\bm{c})$, by optimising $\bm{r}[j]$ in Eq.~\ref{eq:power} and ~\ref{eq_cdi} 
		\If {$|\lambda_1(-(L+C))|*1.001<\Lambda$}
			\State $\bm{r}\gets\{k_1,...,k_{q-1}\}$, $\bm{\eta}\gets\{\eta_1,...,\eta_{q-1}\}$, $\bm{\omega}\gets\{\bm{\omega}_1,...,\bm{\omega}_{q-1}\}$ (i.e. remove input vector
			$\bm{\omega}_j$ and parameters)
		\ElsIf{$|\lambda_1(-(L+C))|*1.001\ge\Lambda$}
		    \State Maximise $|\lambda_1(-(L+C))|$, where $C \gets \text{diag}(\bm{c})$, by optimising $\eta$ where $\bm{\eta}\gets\{\eta,...,\eta\}$ in Eq.~\ref{eq:power} and ~\ref{eq_cdi}
			\State Set $\Lambda \gets |\lambda_1(-(L+C))|$
		\EndIf
 	\EndFor
 	\State Maximise $|\lambda_1(-(L+C))|$ where $C=\text{diag}({\bf p})$ by optimising $\bm{r}$ and $\bm{\eta}$ in Eq.~\ref{eq:power} and~\ref{eq_cdi}
	\State Set $\Lambda \gets |\lambda_1(-(L+C))|$, $q$ is the current number of input vectors ($\bm{\omega}$)
 	\For {$i = 1$ {\bf to} $q$}
         \State $\bm{k}\gets \bm{r}$, then $\bm{r}[i]\gets {\infty}$ (i.e. essentially removing input vector $\bm{\omega}_i$)
 		\State Calculate $|\lambda_1(-(L+C))|$ where $C=\text{diag}({\bf p})$ with $\bm{r}$ and $\bm{\eta}$ in Eq.~\ref{eq:power} and~\ref{eq_cdi}
 		\If {$|\lambda_1(-(L+C))|*1.001<\Lambda$}
 			\State $\bm{r}\gets \bm{k}$
		\ElsIf {$|\lambda_1(-(L+C))|*1.001\ge\Lambda$}
			\State Set $\Lambda = |\lambda_1(-(L+C))|$.
		\EndIf
	\EndFor
	\State $\bm{k}\gets \emptyset$, $\bm{q}\gets \emptyset$, $\bm{w}\gets \emptyset$
	\For{$i = 1$ {\bf to} $q$}
	    \If{$\bm{r}[i] \ll {\infty}$}
	        \State $\bm{k}\gets \{\bm{k},r_i\}$, $\bm{q}\gets \{\bm{q},\eta_i\}$, $\bm{w}\gets \{\bm{w},\bm{\omega}_i\}$
	    \EndIf
	\EndFor
	\State $\bm{r}\gets\bm{k}$, $\bm{\eta}\gets\bm{q}$, $\bm{\omega}\gets\bm{w}$
	\State Maximise $|\lambda_1(-(L+C))|$ where $C=\text{diag}({\bf p})$ by optimising $\bm{r}$ and $\bm{\eta}$ in Eq.~\ref{eq:power} and~\ref{eq_cdi}
%	\EndProcedure
\end{algorithmic}
\end{algorithm} 
A numerical optimiser, employing a sequential quadratic programming method\cite{mathworks2015fminunc}, is used throughout the algorithm to maximise the dominant eigenvalue by optimising the power, ${\bm{\eta} = \{\eta_1,...,\eta_{i}\}}$, and weighting, ${\bm{r} = \{r_1,...,r_{i}\}}$, variables for the $i$ input vectors. The algorithm first optimises the power variables using the power optimisation method for one input vector.  This power is employed for checking if adding any more input vectors and numerically optimising only the new weighting variable will increase the value of $|\lambda_1(-(L+C))|$. If the convergence speed is improved then the new selection of input vectors will have their power variables numerically optimised, before repeating the search for new input vectors and optimising the new weighting variable. Once all input vectors have been checked both the weighting and power variables are optimised numerically. 
   
To check that there are not any redundant input vectors, each vector is removed from the optimisation, starting with the first input vector, to check if that removal increases $|\lambda_1(-(L+C))|$. If removing the vector does not improve the performance then it is reintroduced and the process is repeated for the other community's input vectors. The final combination of input vectors (i.e. communities) are optimised numerically by varying their weighting and power variables to maximise $\lambda_1(-(L+C))$.

\subsection*{Matching Brain Communities}\label{sec:brain_method}
Each brain connectome graph contains 1,827,240 voxels that each represent a 1\,mm\textsuperscript{3} volume of the brain. The centre of each volume (voxel) forms a three dimensional grid with 1\,mm spacing between neighbouring voxels. Each edge in the graph is defined as any two vertices that are connected by at least a single fibre, where an edge of weight 1 would represent a single fibre connection. This results in an undirected network of weighted edges where around half of these voxels have connections in the subjects considered here.

For the large brain connectome graphs, the adjacency matrix was employed, rather than the Laplacian, to identify CDI. This was due to the difficulty that emerged in converging on $\lambda_1=0$ for such large matrices that contained more than one near zero eigenvalue. Both adjacency and Laplacian matrices can be used with the procedure in Algorithm~\ref{Alg:CDI}. The left eigenvectors of these matrices are the same in certain cases, including graphs with a constant outdegree for each vertex. In other cases the eigenvectors vary but the first eigenvector contains all positive entries for both, while the following eigenvectors divide the network in a similar manner.
It should also be noted that due to the undirected nature of the connectome data, the Laplacian matrix's ability to highlight the imbalance between outdegree and indegree is less relevant. In fact, for an undirected Laplacian matrix the first eigenvector is uniform with an imbalance in the indegree and outdegree of vertices required to determine influence from this eigenvector.

The vertices included in influential communities from different graphs were compared to determine similarity. For this similarity comparison, the CDI were reduced in size by only including vertices with a large eigenvector entry. This eigenvector entry threshold was set at 0.01 (i.e. $(v_A)_i>0.01$ where $v_A$ is any of the eigenvectors used in the CDI coordinate system).
The similarity of two graphs was assessed by calculating the number of matching communities shared between both graphs.
This comparison metric for assessing community matches, developed here, considers the shortest distance from all the vertices of one community to the nearest vertex that belongs to another. Vertices were considered overlapping if they are from the same voxel or they are in an adjacent voxel (i.e. maximum overlapping voxel distance $\sqrt{3}\approx1.74$\,mm). The percentage of overlapping vertices are calculated for each community comparison to find the highest percentage overlap between two communities in separate scans. The communities appear to reveal pathways in the brain as depicted in Fig. \ref{fig:ALL_K} {\bf b}. These pathways can sometimes vary in density of vertices and in length, which makes an exact match between two communities unlikely.
Therefore, for a pair of communities to be considered a match their percentage overlap had to exceed a threshold value. The {\it mean number of matching communities} was determined by taking the mean number of matches from a range of threshold values between 50\% and 90\%, at 10\% intervals. 
Note that any community can only be a member of one matching pair, i.e. if a community in one scan 1 matches with multiple communities in another scan only one of those matching pairs would be considered for the number of matching communities.

Finally, it is worth noting that there are always errors in the images produced from MRI scans, even when using the same equipment and procedures, with small errors occurring because of slight changes in image orientation and magnetic field instability \cite{morey2010scan}. The {\it mean number of matching communities} is, therefore, also able to accommodate any small positional errors when detecting overlapping communities.

\subsection*{Spectral Bisection}

The Fiedler vector is associated with the second smallest eigenvalue of the Laplacian matrix but the second eigenvector of the adjacency matrix, used here for the analysis of brain networks, also divides the network in a similar manner. These spectral bisections are completed three times to create 8 communities by first dividing the network according to the second eigenvector, with the sign of its entries determining community division. The second eigenvector is then assessed for both of these communities and more divisions made. For the final bisection of four communities into eight, the second eigenvector was used unless it did not generate two communities with values higher than the threshold used when assessing the {\it mean number of matching communities}, described previously. In the case the next eigenvector that divided the community so that both divisions had values above the threshold is selected. This ensures that all scans have eight eigenvectors with which to compare. .

\bibliography{references}

%\noindent LaTeX formats citations and references automatically using the bibliography records in your .bib file, which you can edit via the project menu. Use the cite command for an inline citation, e.g.  \cite{Hao:gidmaps:2014}.
%
%For data citations of datasets uploaded to e.g. \emph{figshare}, please use the \verb|howpublished| option in the bib entry to specify the platform and the link, as in the \verb|Hao:gidmaps:2014| example in the sample bibliography file.

\section*{Acknowledgements}
This work was supported, in part, by the Engineering and Physical Sciences Research Council [EP/L505080/1].

\section*{Contributions}

R.C., G.P. and M.M. devised the study; R.C. developed the algorithms, performed the analyses, wrote the paper and prepared the figures;  All authors reviewed the manuscript. 

\section*{Competing interests}

The authors declare no competing financial interests. 

\section*{Corresponding author}

Correspondence to Ruaridh Clark.

\end{document}